# Ergodic descriptors of nonergodic stochastic processes

**Madhur Mangalam[1] and Damian G. Kelty-Stephen[2]**

[1]Department of Physical Therapy, Movement and Rehabilitation Sciences, Northeastern University, Boston, MA, USA

[2]Department of Psychology, State University of New York at New Paltz, New Paltz, NY, USA

*ORCIDs:*

Madhur Mangalam (0000-0001-6369-0414)

Damian G. Kelty-Stephen (0000-0001-7332-8486)

*Authors for correspondence:*

Madhur Mangalam

e-mail: m.manglam@northeastern.edu

Damian G. Kelty-Stephen

e-mail: keltystd@newpaltz.edu

**Abstract**

The stochastic processes underlying the growth and stability of biological and psychological systems reveal themselves when far from equilibrium. Far from equilibrium, nonergodicity reigns. Nonergodicity implies that the average outcome for a group/ensemble (i.e., of representative organisms/minds) is not necessarily a reliable estimate of the average outcome for an individual over time. However, the scientific interest in causal inference suggests that we somehow aim at stable estimates of the cause that will generalize to new individuals in the long run. Therefore, the valid analysis must extract an ergodic stationary measure from fluctuating physiological data. So the challenge is to extract statistical estimates that may describe or quantify some of this nonergodicity (i.e., of the raw measured data) without themselves (i.e., the estimates) being nonergodic. We show that traditional linear statistics such as the standard deviation (*SD*), coefficient of variation (*CV*), and root mean square (*RMS*) can show nonstationarity, violating the ergodic assumption. Time series of statistics addressing sequential structure and its potential nonlinearity: fractality and multifractality, change in a time-independent way and fulfill the ergodic assumption. Complementing traditional linear indices with fractal and multifractal indices would empower the study of stochastic far-from-equilibrium biological and psychological dynamics.

**Keywords:** fractal, multifractal, nonequilibrium, stationarity, statistical analysis

## 1. Introduction

The dominant inferential models of the cause and effect in time-varying processes are linear. They assume stable average estimates across independent samples capable of approximating a population within a stable bandwidth of variability. At its core, this assumption is an assumption of ergodicity. Indeed, our customary view of the cause is of an independent factor reliable for producing the same effect on average across a variety of disparate causes. So, our understanding of the cause itself is largely ergodic. A significant question is how well our linearly-modeled, ergodic notion of the cause aligns with our measurements of the systems we wish to explain. A process is ergodic if the ensemble average equals the time average. For example, the Brownian motion of randomly colliding gas molecules in a container is ergodic, entailing that the time a given molecule spends the same time in one half of the container as in the other half's. The assumption of ergodicity is fundamental to the equilibrium statistical mechanics that gave rise to our linear inferential models at the first place [1]. Ergodicity allows replacing dynamical descriptions with simpler probabilistic summary descriptions—essentially eliminating time from the models and supporting our understanding of stable causes [2,3]. A significant challenge then is that systems failing to be ergodic do not satisfy the needs of linear models for inferring cause. For those sciences dealing with non-ergodic systems, a pressing need is to identify or develop statistical descriptors of non-ergodic behavior that will themselves be ergodic. That is, we need ergodic descriptors of non-ergodicity because, short of developing nonlinear models for inferring cause or revamping our notion of the cause to be interdependent and context-sensitively unreliable, non-ergodic behavior will otherwise remain beyond the scope of our scientific explanations.

### 1.1. Why ergodicity matters for biological and psychological sciences?

Biological and psychological sciences are currently coming to grips with how seldom the ergodic assumptions hold. Life and mind are systems that might hover close to an average for brief periods, but beyond a brief convenience sample, the appearance of ergodicity quickly begins to wither away. We might consider a brief caricature to contrast the biological and psychological case with the gas example above. Let us consider individual wealth status over a 100-year life span, particularly its trajectory through a national economy instead of a container: if economic status were ergodic, then we would have to expect that, on multiple, repeated observations, we could find that same individual human inhabiting different regions of the economy in proportion to how many people inhabit those regions over the whole population. Under the ergodic assumption, the individual's economic status would spend 20 years in the blue-collar class, 60 in the lower-middle class, ten years in upper-middle, nine years in upper, and one year in the top one percent. Despite the absurdity of this notion, biological and psychological science have persisted mainly in inferring cause by applying ergodic models assuming that individual-level variability resembles ensemble-level variability and that both are stationary, exhibiting homogeneous mean and variance over time [4–7]. Unfortunately, ergodicity fails in both the theory and in evidence. Theoretically, the form and behavior that distinguish life as an evolving, innovating, and adaptive process are bound towards novelty and divergence. Empirically, measurements of living and thinking systems widely fail tests of ergodicity at all levels of organization: motion of protein molecules with a cell [8–12], hemodynamics and intracellular and extracellular transport of complex media in biological systems, such as cytoplasm and nucleoplasm [13,14], intravascular blood flow [15], coupled networks of heterogeneous neurons [16,17], cognitive processes involved in a variety of perceptuomotor behavior [18–22]. Description by averages is at odds with our expectations of what life is and as noted above, what is at stake is not a mathematical nicety but rather the issue that applying ergodic models to non-ergodic measurement privileges expedience and convention over access to the cause. Ergodic models can make causal inferences from average descriptions of non-ergodic behavior, but there are few reasons to expect that these causal inferences should be correct or replicable.

The valid analysis must extract ergodic stationary descriptors from nonstationarily fluctuating physiological data to avoid making fallacious causal inferences. The problem is that an average descriptor (e.g., arithmetic mean) assuming ergodicity will inherit the measurement's nonergodicity and pass it on to the inferential model, contaminating the inferential validity. A pressing need for valid causal inference is a set of statistical descriptors that may quantify some of this nonergodicity (i.e., of the raw measured data) without themselves (i.e., the descriptors) being nonergodic.

## 1.2. What is ergodicity?

A stochastic process $x(t)$ can support to two types of averaging. First, the finite-ensemble average of the quantity $x$ at a given time $t$ is

$$\langle x(t) \rangle_N = \frac{1}{N} \sum_i^N x_i(t) \tag{1}$$

where $x_i$ is the $i$th of $N$ realizations of $x(t)$ included in the average. Second, the finite-time average of the quantity $x(t)$ is

$$\overline{x_{\Delta t}} = \frac{1}{\Delta t} \int_t^{t+\Delta t} x(s) ds \quad . \tag{2}$$

If $x$ changes at $T = \Delta t / \delta t$ discrete times $t + \delta t$, $t + 2\delta t$, ..., then Eq. (2) becomes

$$\overline{x_{\Delta t}} = \frac{1}{T \delta t} \sum_{\tau+1}^T x(t + \tau \delta t) \quad . \tag{3}$$

An observable $X$ is ergodic if its ensemble average converges to its time average with probability one, such that

$$\lim_{\Delta t \to \infty} \frac{1}{\Delta t} \int_t^{t+\Delta t} X(s) ds = \lim_{N \to \infty} \frac{1}{N} \sum_i^N X_i(t) \quad . \tag{4}$$

Hence, a stochastic process is ergodic if any collection of samples represents the entire process's average statistical properties. Conversely, a stochastic process is nonergodic when its statistics change with time. For example, a gamble can be modeled as in [2]: toss a coin, and for heads, you win 50% of your current wealth of $1; for tails, you lose 40%. The expected wealth across infinite such gambles does not reflect what happens over time (figure 1). For an individual player, wealth trends to zero over time. However, for the aggregate of individuals, wealth trends exponentially upwards over time. A minority of gamblers accrue winning that distorts the average.

## 1.3. The present study

The present study aims to identify a set of statistical descriptors that may quantify the nonergodicity of the raw measured data without themselves being nonergodic, by following up on previous demonstrations that the ergodicity of a noise process may be inversely related to its temporal correlation. The absence of temporal correlation renders additive white Gaussian noise a classic case of ergodicity, but progressively more temporal correlations as in "pink" fractional Gaussian noise can lead ergodicity to break down. This point is particularly relevant to biological and psychological sciences because measurements of physiological, motor, and cognitive variability widely show varying degrees of temporal correlations [23–27]. These temporal correlations can reflect a fractal or even multifractal geometry suggesting interdependence across constituent factors at multiple scales [28]. The multifractal evidence of this interdependence shows that physiological, motor, and cognitive processes may not respect the independence across samples needed for ergodicity. However, fractal and multifractal descriptors of these temporal correlations have repeatedly performed well at predicting and explaining physiological, motor, and cognitive performance in linear inferential models, even including linear causal models like vector autoregression [20,29,30]. That is, fractal and multifractal descriptors bridge a crucial divide: on the one hand, they explicitly encode non-ergodicity of these physiological, motor, and cognitive measurements, and on the other hand, the fractal and multifractal descriptors themselves exhibit sufficient ergodicity across time to support valid causal inference in linear models.

The present work translates these insights about fractal and multifractal descriptors into a theoretical test by examining their ergodicity in simulations of white noise and pink noise. By segmenting white noise and pink noise into nonoverlapping epochs of different sizes, we construct sequences of traditional linear descriptive statistics—standard deviation (*SD*), coefficient of variation (*CV*), and root mean square (*RMS*), and sequences of fractal and multifractal estimates—fractal scaling exponent ($H_{fGn}$), multifractal spectrum width ($\Delta \alpha$), and *t*-statistic comparing spectrum width of the original series to that of its surrogates ($t_{MF}$). We predict no differences in ergodicity between linear and fractal or multifractal descriptors for white noise and shuffled pink noise. In contrast, we predict that for the same epochs over the same pink noise processes, sequences of fractal and multifractal descriptors will be more ergodic than sequences of traditional linear descriptors.

## 2. Methods

### 2.1. Simulating white noise and pink noise

We used 100 simulated 50,000-samples of both white nose and pink noises in MATLAB (Matlab Inc, Natick, MA). White noise was generated with zero mean and unit standard deviation. Pink noise was generated using the function 'pinknoise()'. All series were unsigned and all analysis was conducted on these unsigned values. Using unsigned values is a common practice in fractal and multifractal analysis. A shuffled version of each original white noise and pink noise series was generated for comparison, specifically because ergodicity is about how sequence exemplifies a typical mean trajectory of a sample of realizations. By breaking the sequence, shuffling of all Gaussian noises produces white-noise fluctuations around the mean.

### 2.2. Estimating linear descriptors

We computed the following linear indices over nonoverlapping epochs of 250, 500, 100, and 2000 samples—amounting to 200, 100, 50, and 25 epochs, respectively—for the original version (i.e., unshuffled) and a shuffled version (i.e., a version with the temporal information destroyed) of each pink and white noise series:

Standard deviation (*SD*), that is,

$$\sqrt{\frac{1}{N}\sum_{i=1}^{N}(x_i-\mu)^2}, \tag{5}$$

where $N$ is sample size (i.e., 250, 500, 100, or 2000), $x_i$ is $i$th individual value, and $\mu$ is sample mean.

Coefficient of variation (*CV*), that is,

$$\frac{SD}{\mu}, \tag{6}$$

where $\mu$ is sample mean.

Root mean square (*RMS*), that is,

$$\sqrt{\frac{1}{N}\sum_{i=1}^{N}|x_i|^2}, \tag{7}$$

where where $N$ is sample size, $x_i$ is $i$th individual value, and $\mu$ is sample mean.

### 2.3. Fractal and multifractal descriptors

### 2.3.1. Accessing fractality using detrended fluctuation analysis

Detrended fluctuation analysis (DFA) computes Hurst exponent $H_{fGn}$ quantifying temporal correlations in white noise and pink noise series [31,32] using the first-order integration of $N$-length time series $x(t)$:

$$y(t)=\sum_{i=1}^{N}x(t)-\overline{x(t)}, \tag{8}$$

where $\overline{x(t)}$ is the time-series grand mean. It computes root mean square (RMS; i.e., averaging the residuals) for each linear trend $y_n(t)$ fit to nonoverlapping $n$-length bins ()to build fluctuation function $f()$:

$$f(N)=\sqrt{(1/N)\sum_{i=1}^{N}(x(t)-\overline{x(t)})^2}, \tag{9}$$

for $n<N/4$. On standard scales, $f(N)$ is a power law:

$$f(N)\sim n^H, \tag{10}$$

where $H_{fGn}$ is the scaling exponent estimable using logarithmic transformation:

$$\log f(N)=H_{fGN}\log(n). \tag{11}$$

Higher $H_{fGn}$ corresponds to stronger temporal correlations. We computed $H_{fGn}$ over 250-, 500-, 100-, and 2000-sample epochs for the original and shuffled unsigned white noise and pink noise series.

### 2.3.2. Multifractal analysis

### 2.3.2.1. Assessing multifractal nonlinearity using the direct-estimation of singularity spectrum

Chhabra and Jensen's [33] direct method estimated multifractal spectrum widths $\Delta\alpha$ for each 250-, 500-, 100-, and 2000-samples epoch for the original and shuffled unsigned pink noise and white noise series. This method samples series $u(t)$ at progressively larger scales using the proportion of signal $P_i(L)$ falling within the $i^{th}$ bin of scale $L$ is

$$P_i(L) = \frac{\sum_{k=(i-1)L+1}^{iL} u(k)}{\sum u(t)} \tag{12}$$

As $L$ increases, $P_i(L)$ represents progressively larger proportion of $u(t)$,

$$P(L) \propto L^\alpha \tag{13}$$

suggesting growth of proportion according to one "singularity" strength $\alpha$ [34]. $P(L)$ exhibits multifractal dynamics when it grows heterogeneously across time scales $L$ according to multiple singularity strengths, such that

$$P_i(L) \propto L^{\alpha_i} \tag{14}$$

whereby each $i^{th}$ bin may show a distinct relationship of $P(L)$ with $L$. The width of this singularity spectrum, $\Delta\alpha$ ($\alpha_{max} - \alpha_{min}$), indicates the heterogeneity of these relationships [35,36].

Chhabra and Jensen's method [33] estimates $P(L)$ for $N_L$ nonoverlapping bins of $L$-sizes and transforms them into a "mass" $\mu(q)$ using a $q$ parameter emphasizing higher or lower $P(L)$ for $q>1$ and $q<1$, respectively, as follows

$$\mu_i(q, L) = \frac{[P_i(L)]^q}{\sum_{i=1}^{N_L} [P_i(L)]^q} . \tag{15}$$

$\alpha(q)$ is the singularity for mass $\mu$-weighted $P(L)$ estimated by

$$\alpha(q) = -\lim_{L \to \infty} \frac{1}{\ln L} \sum_{i=1}^{N} \mu_i(q, L) \ln P_i(L)$$

$$= \lim_{L \to 0} \frac{1}{\ln L} \sum_{i=1}^{N} \mu_i(q, L) \ln P_i(L) . \tag{16}$$

Each estimated value of $\alpha(q)$ belongs to the multifractal spectrum only when the Shannon entropy of $\mu(q,l)$ scales with $L$ according to the Hausdorff dimension $f(q)$, where

$$f(q) = -\lim_{L \to \infty} \frac{1}{\ln L} \sum_{i=1}^{N} \mu_i(q, L) \ln \mu_i(q, L)$$

$$= \lim_{L \to 0} \frac{1}{\ln L} \sum_{i=1}^{N} \mu_i(q, L) \ln \mu_i(q, L) .$$
(17)

For values of $q$ yielding a strong relationship between Eqs. (16 & 17)—in this study, $r > 0.995$, the parametric curve $(\alpha(q), f(q))$ or $(\alpha, f(\alpha))$ constitutes the multifractal spectrum. We obtained the multifractal-spectrum width $\Delta\alpha$ over nonoverlapping epochs of 250, 500, 100, and 2000 samples for the original and shuffled versions of each white noise and pink noise series.

### 2.3.2.2. Surrogate testing using Iterated Amplitude Adjusted Fourier Transformation (IAAFT) generated $t_{MF}$

To identify whether nonzero $\Delta\alpha$ reflected multifractality due to nonlinear interactions across timescales, $\Delta\alpha$ for the original and a shuffled version of each pink and white noise series was compared to $\Delta\alpha$ for 32 IAAFT surrogates [28,37]. IAAFT randomizes original values time-symmetrically around the autoregressive structure, generating surrogates that randomized phase ordering of the series' spectral amplitudes while preserving linear temporal correlations. The one-sample *t*-statistic (henceforth, $t_{MF}$) takes the subtractive difference between $\Delta\alpha$ for the original series and that for the 32 surrogates, dividing by the standard error of the spectrum width for the surrogates. We obtained $t_{MF}$ computed over nonoverlapping epochs of 250, 500, 100, and 2000 samples for the original and shuffled versions of each white noise and pink noise series.

## 2.5. Estimating ergodicity breaking in raw and SD, CV, RMS, $H_{fGn}$, Δa, and $t_{MF}$ series

Ergodicity can be quantified using a dimensionless statistic of ergodicity breaking $E_B$ — also known as the Thirumalai-Mountain metric [38], computed by subtracting squared the total-sample variance from the average squared subsample variance and dividing the resultant by the total-sample squared variance.

$$E_B(x(t)) = \frac{(\langle [\delta^2(x(t))]^2 \rangle - \langle \overline{\delta^2}(x(t)) \rangle)}{\langle \overline{\delta^2}(x(t)) \rangle^2} \tag{18}$$

Rapid convergence of $E_B$ for progressively larger samples, that is, $E_B \to 0$ as $t \to \infty$ implies ergodicity. Slower convergence indicates weaker ergodicity, and no convergence indicates nonergodicity [39]. $E_B$ thus serves a simple way to test whether a given statistical descriptor fulfills ergodic assumptions or breaks ergodicity and compare nonergodicity of one statistical descriptor to that of the other. We computed the ergodicity breaking factor $E_B$ for each original and shuffled raw unsigned white noise and pink noise series, for lag Δ = 2 samples, and for each SD, CV, RMS, $H_{fGn}$, Δa, and $t_{MF}$ series computed over nonoverlapping epochs of 250, 500, 100, and 2000 samples each for the original and shuffled unsigned white noise and pink noise, for Δ = 2 epochs.

## 3. Results

### 3.1. Pink noise shows nonstationarity, violating the ergodic assumption

The unsigned pink-noise series shows higher variability than the unsigned white-noise signal at all ensemble sizes. Even a visual inspection of figures 2*a* and 2*b* is enough to observe this distinction, reflecting potential differences in ergodicity: unsigned white noise is relatively more likely than unsigned pink noise to return and, over a larger ensemble, converge towards the mean, and hence, is more ergodic than pink noise. This nonergodic behavior of pink noise is evident in the $E_B$ vs. epochs curves, which coincide almost completely for the original and shuffled unsigned white noise but do not coincide for the original and shuffled unsigned pink noise (figures 2*c* and 2*d*). Indeed, the original unsigned pink noise shows ergodicity breaking with a shallower $E_B$ curve, although the $E_B$ vs. epochs curve for the shuffled unsigned pink noise was no distinct than the $E_B$ vs. epochs curves for the original and shuffled unsigned white noise. So overall, pink noise shows nonstationarity, violating the ergodic assumption.

### 3.2. Linear statistics such as *SD*, *CV*, and *RMS* show nonstationarity, violating the ergodic assumption

We first examine the ergodicity breaking behavior of *SD*, *CV*, and *RMS* series using the Thirumalai-Mountain method, with each of these statistical descriptors computed across

50 epochs of 1000-samples each. No difference exists between *SD* series for the original and shuffled unsigned white noise (figure 3*a*), but *SD* series for shuffled unsigned pink noise is marginally lower than the *SD* series for the original unsigned pink noise (figure 3*b*), reflecting that variability grows faster in unsigned pink noise than unsigned white noise. Although *SD* series for the original and shuffled unsigned white noise did not differ in their ergodicity breaking behavior (figure 3*c*), *SD* series for the original unsigned pink noise showed stronger ergodicity breaking than *SD* series for the shuffled unsigned pink noise (figure 3*d*).

No difference exists between *CV* and *RMS* series for the original and shuffled unsigned white noise (figures 4*a* and 5*a*), but *CV* and *RMS* series for shuffled unsigned pink noise is marginally lower than *CV* and *RMS* series for the original unsigned pink noise (figures 4*b* and 5*b*). As opposed to *SD*, which was higher for pink noise than white noise, *CV* and *RMS* was comparable for white noise and pink noise, reflecting that variability grows faster in unsigned pink noise than unsigned white noise, bit this growth is proportional to the growth of the mean. Although *CV* and *RMS* series for the original and shuffled unsigned white noise did not differ in their ergodicity breaking behavior (figures 4*c* and 5*c*), *CV* and *RMS* series for the original unsigned pink noise showed stronger ergodicity breaking than their shuffled counterparts (figures 4*d* and 5*d*).

In summary, *RMS* series behave exactly as *CV* series, although the two show some differences from *SD* series in its ergodicity breaking behavior. Nonetheless, *SD*, *CV*, and *RMS* series all showed ergodicity breaking for the original unsigned pink noise, demonstrating that computing these statistical descriptors across 50 epochs of 1000-samples each did not change the ergodicity breaking behavior of the original unsigned pink noise. In other words, *SD*, *CV*, and *RMS* fail to encode the nonergodicity inherent to pink noise, which occurs most commonly in biological and psychological processes [40–42].

### 3.3. Fractal and multifractal statistics such as $H_{fGn}$, $\Delta a$, and $t_{MF}$ change in a time-independent way and fulfill the ergodic assumption

We now examine the ergodicity breaking behavior of $H_{fGn}$, $\Delta a$, and $t_{MF}$ series using the Thirumalai-Mountain method, with each of these statistical descriptors computed across 50 epochs of 1000-samples each. No difference exists between $H_{fGn}$ series for the original and shuffled unsigned white noise (figure 6*a*), but $H_{fGn}$ series for the original unsigned pink noise show higher and, for larger ensembles, more consistently higher values than for the original unsigned white noise (figure 6*b*). Notably, $H_{fGn}$ series for the original unsigned pink noise shows none of the ergodicity breaking shown by the raw original unsigned pink noise (figure 6*d*), and the $E_B$ vs. epochs curves were no distinct than those for $H_{fGn}$ series for the original and shuffled unsigned white noise (figure 6*d*).

Much like for $H_{fGn}$ series, $\Delta a$ series do not differ between the original and shuffled unsigned white noise, and they take greater values for the original unsigned pink noise than for the original unsigned white noise, showing a consistent differences for larger ensembles (figures 7*a* and 7*b*). On the contrary, $E_B$ vs. epochs curves scarcely differ between $\Delta a$ series for the original and shuffled unsigned white noise and for the original and shuffled unsigned pink noise (figures 7*c* and 7*d*). $t_{MF}$ series behaved similarly. $t_{MF}$ series do not differ between the original and shuffled unsigned white noise, and they take greater values for the original unsigned pink noise than for the original unsigned white noise (figures 8*a* and 8*b*). $E_B$ vs. epochs curves scarcely differ between $t_{MF}$ series for the original and shuffled unsigned white noise and for the original and shuffled unsigned pink noise (figures 8*c* and 8*d*). So $E_B$ vs. epochs curves corresponding to $\Delta a$ and $t_{MF}$ series for the original and shuffled unsigned pink noise practically overlap, indicating no ergodicity breaking in both these statistical descriptors.

In summary, fractal and multifractal statistics such as $H_{fGn}$, $\Delta\alpha$, and $t_{MF}$ series do not show ergodicity breaking as linear statistics such as *SD*, *CV*, and *RMS* show. These statistics change in a time-independent way and full ergodic assumption—$H_{fGn}$, $\Delta\alpha$, and $t_{MF}$ each encode the nonergodicity inherent to pink noise. Specifically, these fractal and multifractal statistics themselves encode the degree and type of nonergodicity in statistical descriptors.

### 3.4. Effects of lengths of epochs on ergodicity breaking

To investigate whether the observed ergodicity breaking for different linear and nonlinear statistical descriptors depend on epoch size (250, 500, 1000, and 2000 samples, that is, statistical descriptors computer over 200, 100, 50, and 25 epochs), we examine the $E_B$ vs. epochs curves obtained from the Thirumalai-Mountain method for epochs of different sizes.

The $E_B$ vs. epochs curves for the original and shuffled unsigned white noise observed in figures 3 through 6 are largely insensitive to the setting of epoch size, for *SD* series (figure 9*a*), *CV* series (figure 9*b*), and *RMS* series (figure 9*c*). In contrast, *CV* and *RMS* series for the original unsigned pink noise differ from that for the shuffled unsigned pink noise over all epoch sizes (figures 9*d* and 9*f*), although no such difference is observed for *SD* series (figure 9*b*), as suggested in figure 3. However, $E_B$ vs. epochs shows a shallower slope for *SD* series for the original unsigned pink noise than for *SD* series for the shuffled unsigned pink noise for the smallest epochs, suggesting that *SD* may behave ergodically but only for relatively large epochs, and in the algorithmic attempt to rescale *SD* by the mean, *CV*, and *RMS* retain the same failure of ergodicity at all epoch sizes.

The $E_B$ vs. epochs curves for the original and shuffled unsigned white noise and pink noise observed in figures 6 through 8 are largely insensitive to the setting of epoch size, for $H_{fGn}$ series (figures 10*a* and 10*b*), $\Delta\alpha$ series (figures 10*c* and *d*), and $t_{MF}$ series (figures 10*e* and *f*). Hence, while the linear descriptors based on quantifying the second moment show ergodicity breaking, nonlinear descriptors quantifying fractal and multifractal fluctuations showed ergodic behavior that did not depend on epoch size used for the analysis. This conclusion is further strengthened by the observation that $E_B$ vs. epochs curves for $\Delta\alpha$ and $t_{MF}$ both become steeper for the original unsigned pink noise, leaving them steeper than and as steep as, respectively, compared to $E_B$ vs. epochs curves for their shuffled counterparts.

### 4. Discussion

We proposed the present simulation to investigate whether assuming ergodicity will leave our statistical descriptions as relatively ergodic or non-ergodic as the measurement series themselves. As stated above, we want statistical descriptions to be ergodic because all dominant causal inference operates on the assumption of ergodicity. Specifically, we predicted that the traditional linear descriptive statistic would inherit a trace of the raw measurement series' relative ergodicity (or nonergodicity). Secondarily, we predicted fractal and multifractal descriptors designed with assumptions open to the nonlinear interdependence of fluctuations across timescales. Previous work found that pink noise is less ergodic than white noise. The present work replicated that fact and provided support for our predictions. We show that, in particular, for ergodic white noise, the sequences of linear statistical descriptors (*SD*, *CV*, and *RMS*) and the sequences of fractal and multifractal descriptors ($H_{\text{fgn}}$, $\Delta a$, and $t_{\text{MF}}$) applied to moving epochs of the series both exhibited comparable ergodicity. Furthermore, fractal and multifractal descriptors of pink noise were comparably ergodic to fractal and multifractal descriptors of pink noise shuffled to constitute white noise. However, linear statistical descriptors were less ergodic for the original pink noise than for shuffled pink noise constituting white noise.

The ergodicity of linear descriptors was not uniformly sensitive to the ergodicity of the source series. For instance, *SD* for the shorter sequences of longer epochs was comparably ergodic for pink noise and shuffled pink noise. It was only for shorter epochs that the sequences of *SD* became less ergodic. This point is unsurprising because *SD* grows according to the square root of sample size, increasing very quickly with over smaller sample sizes but stabilizing or growing more slowly over larger sample sizes. However, the failure of *CV* and *RMS* to be as ergodic for the original unsigned pink noise as for the shuffled unsigned pink noise suggests that the removal of the sample mean, for example, as in the algorithm for *SD* is crucial for any hope of ergodic statistical description. *CV* and *RMS* both bear the imprint of the mean, whether in dividing by the mean (i.e., for *CV*) or in the omitting to subtract the mean (i.e., for *RMS*). Carrying an algorithmic trace of the mean value leads these two descriptors to carry on the non-ergodicity of the candidate series.

Conventional discourse and statistical practice could make these results challenging to understand. For one, it may seem counterintuitive that the *SD* of a non-ergodic series could be stable. After all, whatever we may say about using the mean for calculating *RMS* and *CV*, *SD* depends no less on the stability of a mean for its calculation. However, what we observe in this result is a signature of the stationarity of pink noise—pink noise offers an intriguing case of stationarity which can nonetheless exhibit nonergodicity for stronger temporal correlations [39]. The subtraction of the mean in the calculation of *SD* might thus remove nonergodicity in pink noise resulting from any temporal correlations.

Meanwhile, these results also cast what may be a startling light on statistical descriptors that have elsewhere seemed strategic. For instance, the use of *RMS* has long served neuroscience, kinesiology, and engineering attempts to quantify a bulk measure that covers both magnitude and variability [43–47]. However, this strategy risks conflating two parameters of the linear model, which should, under ideal circumstances, be independent [48]. Additionally, the *RMS* quantification of variability without removing the mean does appear to make this workhorse of a statistical descriptor unreliable for causal models assuming ergodicity. Similarly, using *CV* is an elegant strategy to resolve the potential violation of homoscedasticity in which variance increases with the mean: dividing *SD* by the mean seems to solve this problem at first glance and offer a closer approximation of the needed homoscedasticity for linear modeling. Notably, the present results suggest that even dividing by the mean is enough to make this descriptor nonergodic.

The encouraging news here is that *SD* need not stand alone in causal models assuming ergodicity. A critical avenue that appears ahead is the possibility of modeling causality in a nonergodic system using *SD* hand in hand with fractal and multifractal descriptors. Temporal correlations are at least one statistical element triggering the evidence of ergodicity breaking—and they are frequent features of biological and psychological processes [22,49–51]. The nonergodicity of *RMS* and *CV* offer the caveat that turning a blind eye to these failures of independence could carry the nonergodicity forward and make causal modeling impossible. However, the ergodicity of fractal and multifractal descriptors suggests that confronting and quantifying these dependencies across time and scale is sooner a means to modeling cause in ergodic terms. Building causal models in biological and psychological that include *SD* alongside fractal and multifractal descriptors has shown early promise, e.g., for explaining changes in *SD* of perceptuomotor performance [29]. We expect that the biological and psychological sciences will profit from broadening its scope to include the possibility that the fractal and multifractal dynamics contributing to non-ergodicity may constitute an important class of causal factors.

**Funding.** We received no funding for this work.

**Competing interests.** We declare we have no competing interests.

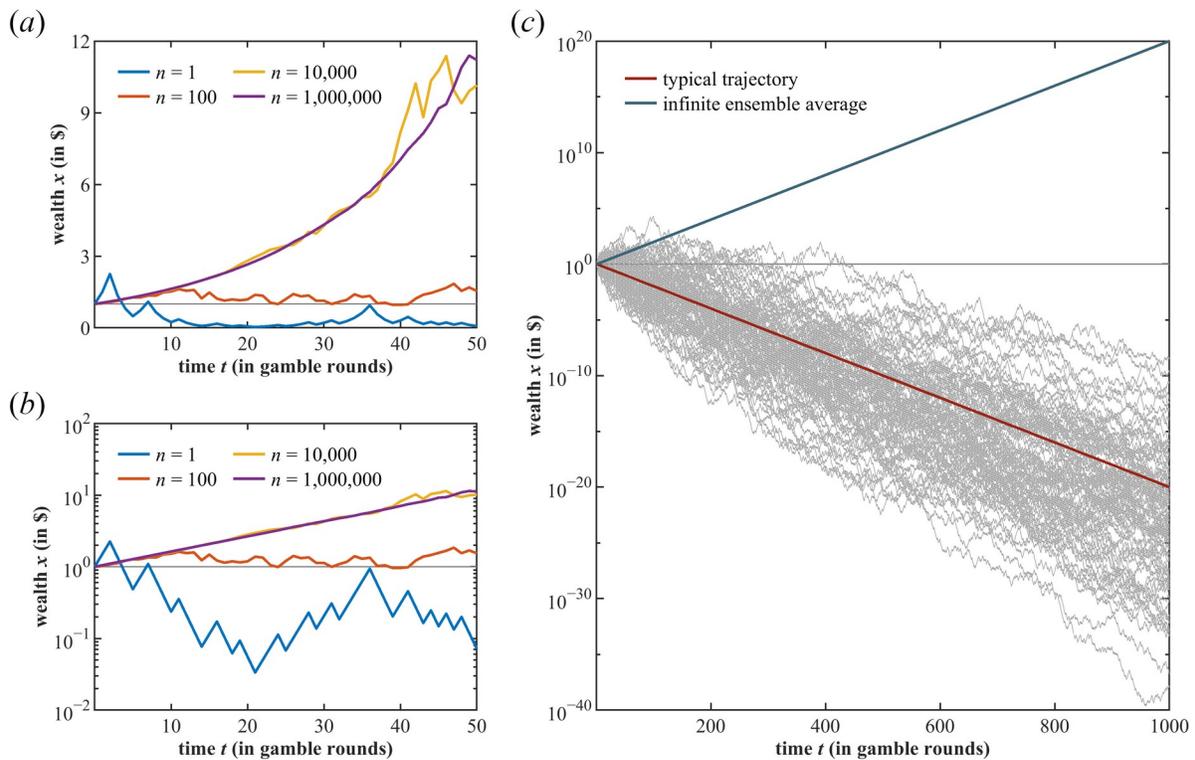

**Figure 1.** Repeated gambling is a nonergodic process. The gamble is modeled as the following: toss a coin, and for heads you win 50% of your current wealth of $1, for tails you lose 40% of your current wealth. (*a*) A gambler goes almost bust over 50 rounds (blue line; but never goes completely bust—since a loss is 40% of the current wealth). The average wealth of 100 concurrent players shows no trend (red line). Simulating 10,000 concurrent players starts showing a trend (yellow line). On average, this group harvest a nice ~10x profit. Simulating a million concurrent players makes this trend more apparent (purple line). (*b*) A logarithmic *y*-axis shows the simulation in (*b*) more clearly. (*c*) The average wealth computed over an infinite ensemble of individuals (blue line) does not reflect what happens to the wealth of a typical individual over time (red line). The illustration consists of 100 trajectories of 1000 repetitions each. Adapted from [2].

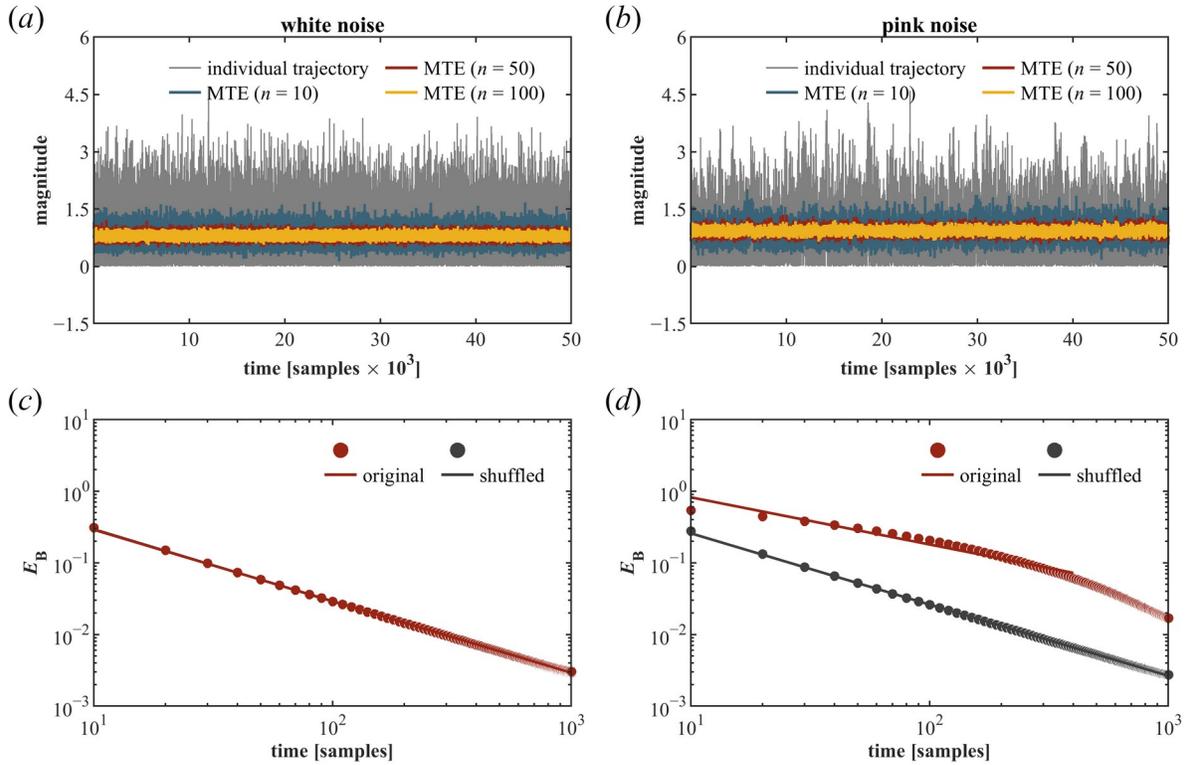

**Figure 2.** Ergodicity breaking in unsigned white noise and pink noise series. (*a*) A simulated unsigned white-noise series and multiple trajectory ensemble (MTE) average over 10, 50, and 100 simulations. (*b*) A simulated unsigned pink-noise series and MTE average over 10, 50, and 100 simulations. A noteworthy difference between (*a*) and (*b*) is that the unsigned pink-noise series shows marginally more variability than the unsigned white-noise signal at all ensemble sizes, from the individual series to the ensembles of 10, 50, and 100 series. So, even just visual inspection provides an early sense of a potential difference in ergodicity: unsigned white noise appears relatively more likely than unsigned pink noise to return and, over a larger ensemble, converge towards the mean. (*c*) and (*d*) show how the ergodicity-breaking parameter ($E_B$; for lag $\Delta$ = 2 samples) differs between the original unsigned white- and pink-noise series and their shuffled versions. This comparison is critical because ergodicity is fundamentally about the sequence—how the sequence of a series exemplifies a typical mean trajectory of a sample of comparable systems. Shuffling breaks the sequence and guarantees that the series' trajectory fluctuates around the mean. (*c*) $E_B$ vs. time *t* for unsigned white-noise series (mean±s.d. $H_{fGn}$ = 0.50±0.03) and shuffled series ($H_{fGn}$ = 0.49±0.03). Because unsigned white noise has no temporal correlations, shuffling an unsigned white-noise series produces another series of unsigned white noise. Hence, the two $E_B$ vs. *t* curves coincide almost completely, with the grey shuffled-series' curve scarcely visible with greater difference in the right half. (*d*) $E_B$ vs. *t* for unsigned pink-noise series ($H_{fGn}$ = 0.69±0.05) and shuffled series ($H_{fGn}$ = 0.50±0.04). Notably, shuffling an unsigned pink-noise series produces an unsigned white-noise series, illustrated by the fact that the grey curve for shuffled unsigned pink-noise series resembles both curves in (*c*). As $H_{fGn} \to 1$, we get nonergodic behavior as indicated by the higher and, for all but the largest sample size, shallower $E_B$ curve [39].

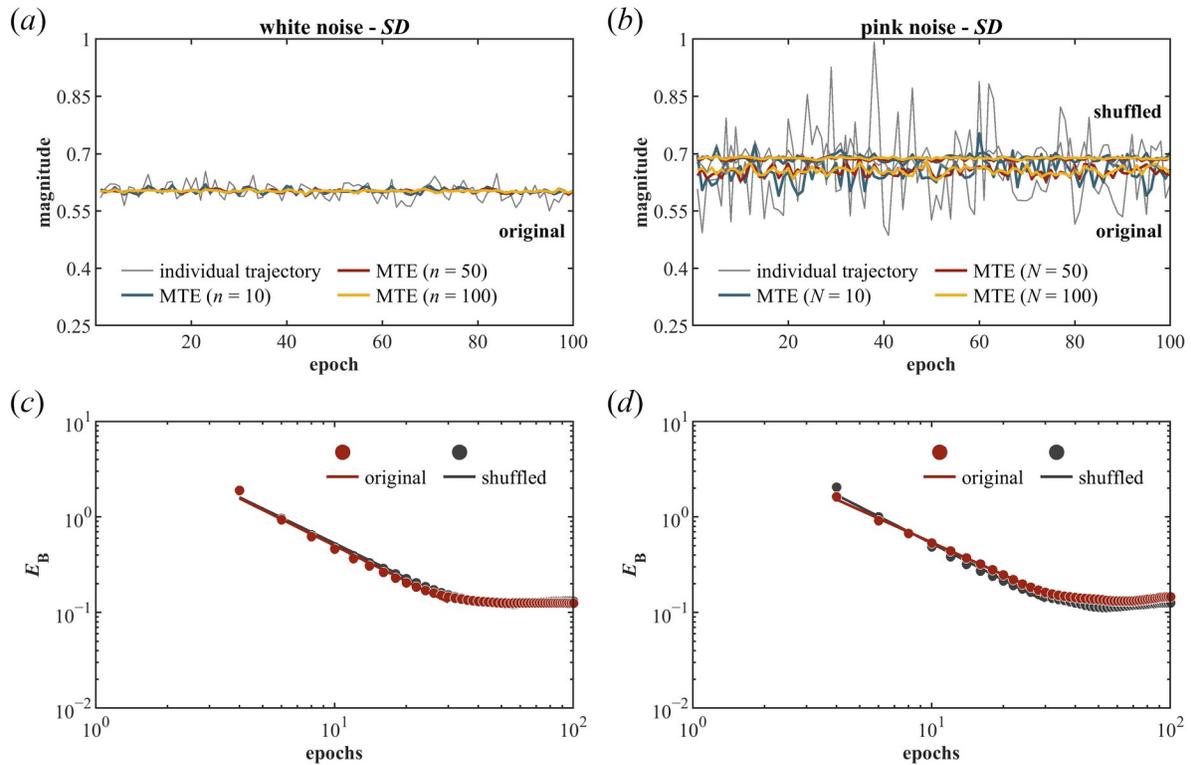

**Figure 3.** Ergodicity breaking in *SD* series (*SD* calculated across 500-samples epochs). *SD* series and MTE average of SD over 10, 50, and 100 simulations for unsigned white noise (*a*; shuffled series not pictured because of overlap) and unsigned pink noise (*b*). *SD* series for shuffled unsigned pink noise is marginally higher than the *SD* series for the original unsigned pink noise, but no difference exists between *SD* series for the original and shuffled unsigned white noise. The greater *SD* in (*b*) than (*a*) reflects that variability can grow faster in unsigned pink noise than unsigned white noise. $E_B$ vs. epochs (for lag Δ = 2 epochs) for *SD* series for the original and shuffled unsigned white noise (*c*) and *SD* series for the original and shuffled unsigned pink noise (*d*). $E_B$ for *SD* series decreases steadily with *s* and then levels off at larger values at *s*. A relatively slower decay in $E_B$ with *s* is observed for *SD* series for the original than shuffled unsigned pink noise, but no such distinction is observed between *SD* series for the original and shuffled unsigned white noise.

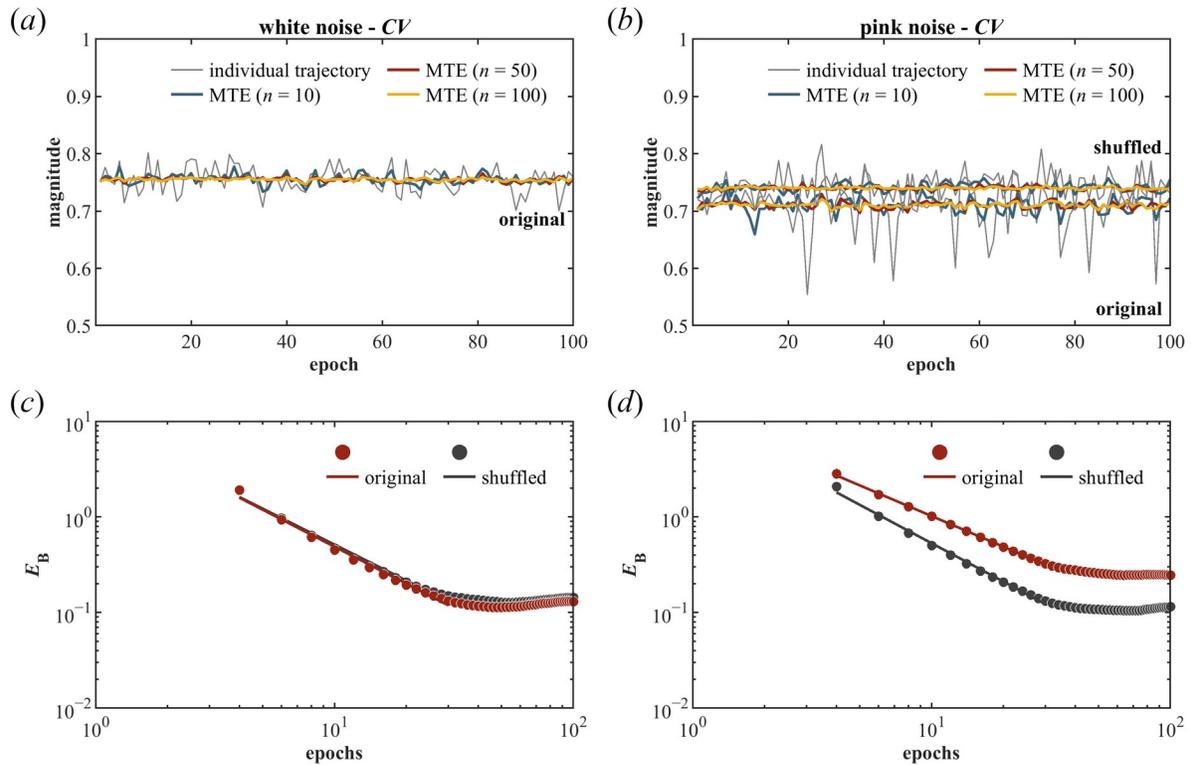

**Figure 4.** Ergodicity breaking in *CV* series (*CV* calculated across 500-samples epochs). *CV* series and MTE average of *CV* over 10, 50, and 100 simulations for unsigned white noise (*a*; shuffled series not pictured because of overlap) and unsigned pink noise (*b*). *CV* series for shuffled unsigned pink noise is marginally higher than the *CV* series for the original unsigned pink noise, but no difference exists between *CV* series for the original and shuffled unsigned white noise. The similar *CV* in (*a*) and (*b*) reflects that, although variability grows faster in unsigned pink noise than unsigned white noise, this growth of variability in unsigned pink noise is proportional to growth of the mean. $E_B$ vs. epochs (for lag $\Delta = 2$ epochs) for *CV* series for the original and shuffled unsigned white noise (*c*) and *CV* series for the original and shuffled unsigned pink noise (*d*). $E_B$ for *CV* series decreases steadily with sample size and then levels off at larger sample sizes. A relatively slower decay in $E_B$ with *s* is observed for *CV* series for the original than shuffled unsigned pink noise, but no such distinction is observed between *CV* series for the original and shuffled unsigned white noise. This slower decay simply suggests that *CV* for unsigned pink noise is less ergodic than *CV* for unsigned white noise.

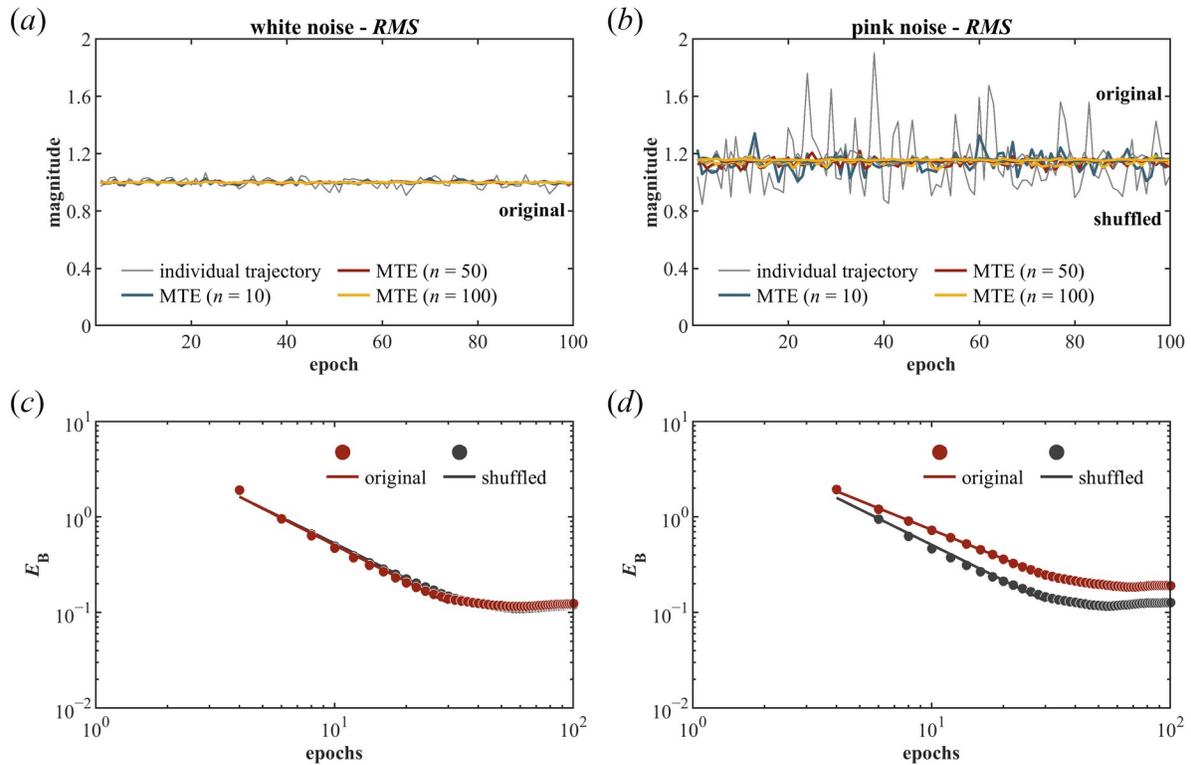

**Figure 5.** Ergodicity breaking in *RMS* series (*RMS* calculated across 500-samples epochs). *RMS* series and MTE average of RMS over 10, 50, and 100 simulations for unsigned white noise (*a*; shuffled series not pictured because of overlap) and unsigned pink noise (*b*). Although *RMS* series has consistently higher values for unsigned pink noise than unsigned white noise, no difference exists between *RMS* series for the original and shuffled unsigned white noise or pink noise. $E_B$ vs. epochs (for lag $\Delta = 2$ epochs) for *RMS* series for the original and shuffled unsigned white noise (*c*) and *RMS* series for the original and shuffled unsigned pink noise (*d*). A relatively slower decay in $E_B$ with *s* is observed for *RMS* series for the original than shuffled unsigned pink noise, but no such distinction is observed between *RMS* series for the original and shuffled unsigned white noise. This slower decay simply suggests that *RMS* for unsigned pink noise is less ergodic than *RMS* for unsigned white noise.

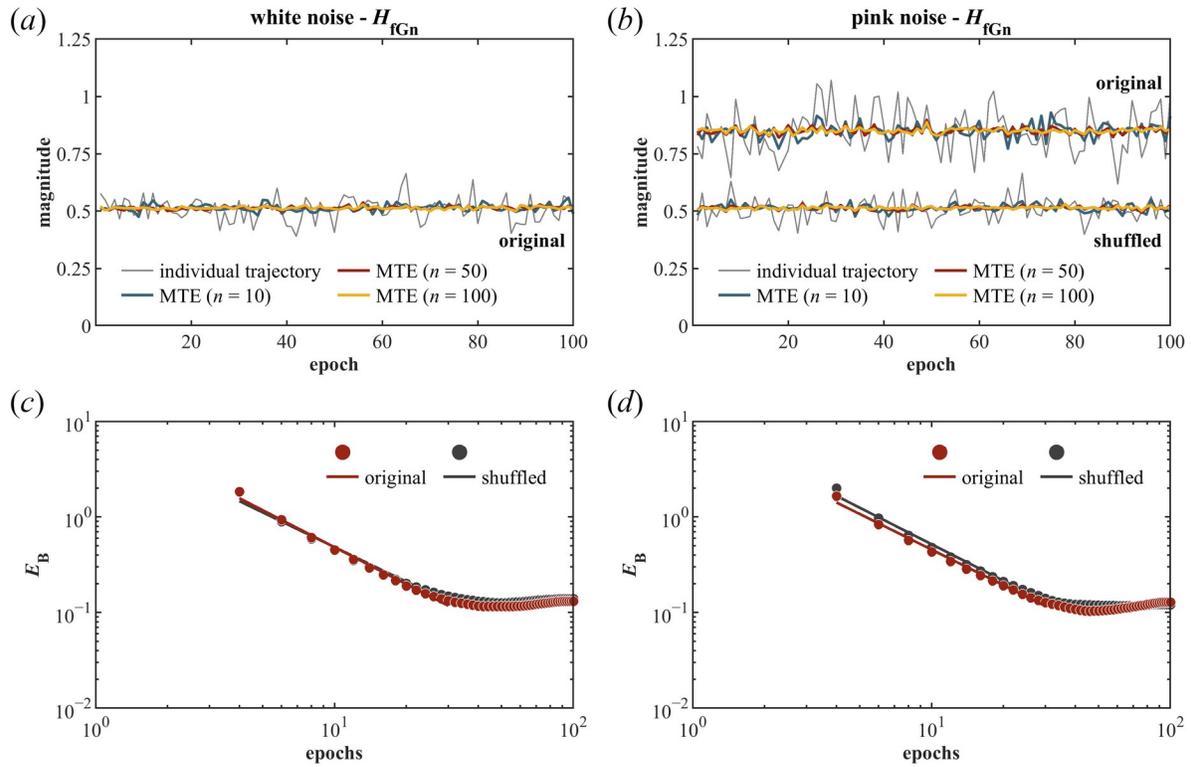

**Figure 6.** Ergodicity breaking in $H_{fGn}$ series ($H_{fGn}$ calculated across 500-samples epochs). $H_{fGn}$ series and MTE average of $H_{fGn}$ over 10, 50, and 100 simulations for unsigned white noise (*a*; shuffled series not pictured because of overlap) and unsigned pink noise (*b*). Although $H_{fGn}$ series for the original and shuffled unsigned white noise show no difference (*a*), $H_{fGn}$ series for the original unsigned pink noise show higher and, for larger ensembles, more consistently higher values than for the original unsigned white noise. Indeed, as noted in figure 2, $H_{fGn}$ series for the shuffled unsigned pink noise is indistinguishable from $H_{fGn}$ series for both the original and shuffled unsigned white noise. $E_B$ vs. epochs (for lag $\Delta = 2$ epochs) for $H_{fGn}$ series for unsigned white noise (*c*) and $H_{fGn}$ series for absolute-value pink noise (*d*). What emerges is that $H_{fGn}$ series shows none of the ergodicity breaking that the raw series for the original unsigned pink noise did in figure 2. The decay in $E_B$ vs. epochs only differs slightly between $H_{fGn}$ series for original and shuffled unsigned white noise at the largest sample sizes. It shows almost no difference between $H_{fGn}$ series for the original and shuffled unsigned pink noise—and whatever difference exists indicates that $H_{fGn}$ series are more ergodic (i.e., exhibiting lower $E_B$) for the original unsigned pink noise than for the shuffled unsigned pink noise.

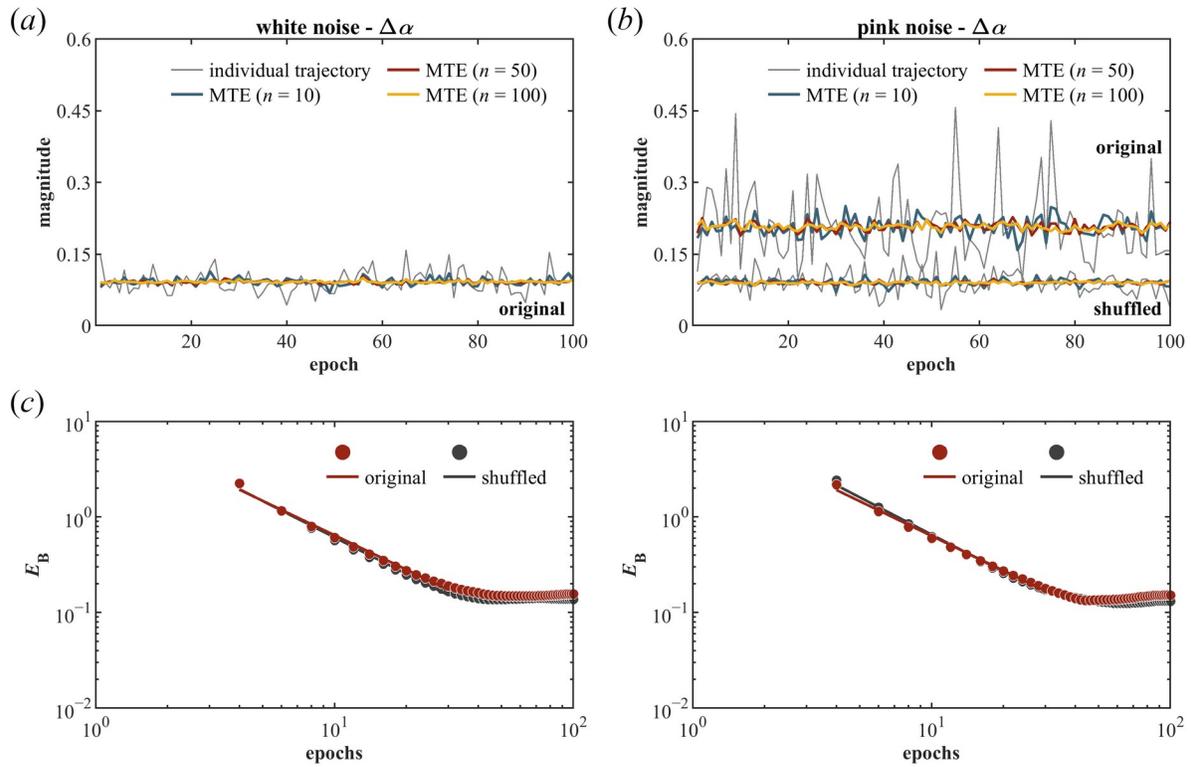

**Figure 7.** Ergodicity breaking in Δ$a$ series (Δ$a$ calculated across 500-samples epochs). Δ$a$ series and MTE average of Δ$a$ over 10, 50, and 100 simulations for unsigned white noise (*a*; shuffled series not pictured because of overlap) and unsigned pink noise (*b*). Much like for $H_{fGn}$ series in figures 6*a* and 6*b*, Δ$a$ series do not differ between the original and shuffled unsigned white noise, and they take greater values for the original unsigned pink noise than for the original unsigned white noise, showing a consistent differences for larger ensembles. $E_B$ vs. epochs (for lag Δ = 2 epochs) scarcely differs between Δ$a$ series for the original and shuffled unsigned white noise (*c*) and Δ$a$ series for the original and shuffled unsigned pink noise (*d*). All cases show a in $E_B$ with *s* which slows only at the largest sample sizes. These $E_B$ vs. epochs curves practically overlap, with Δ$a$ series for the original unsigned white noise showing slightly less ergodicity (i.e., higher $E_B$; *c*) and with Δ$a$ series for original unsigned pink noise showing slightly more ergodicity (i.e., lower $E_B$; *d*) than Δ$a$ series for their respective shuffled unsigned series.

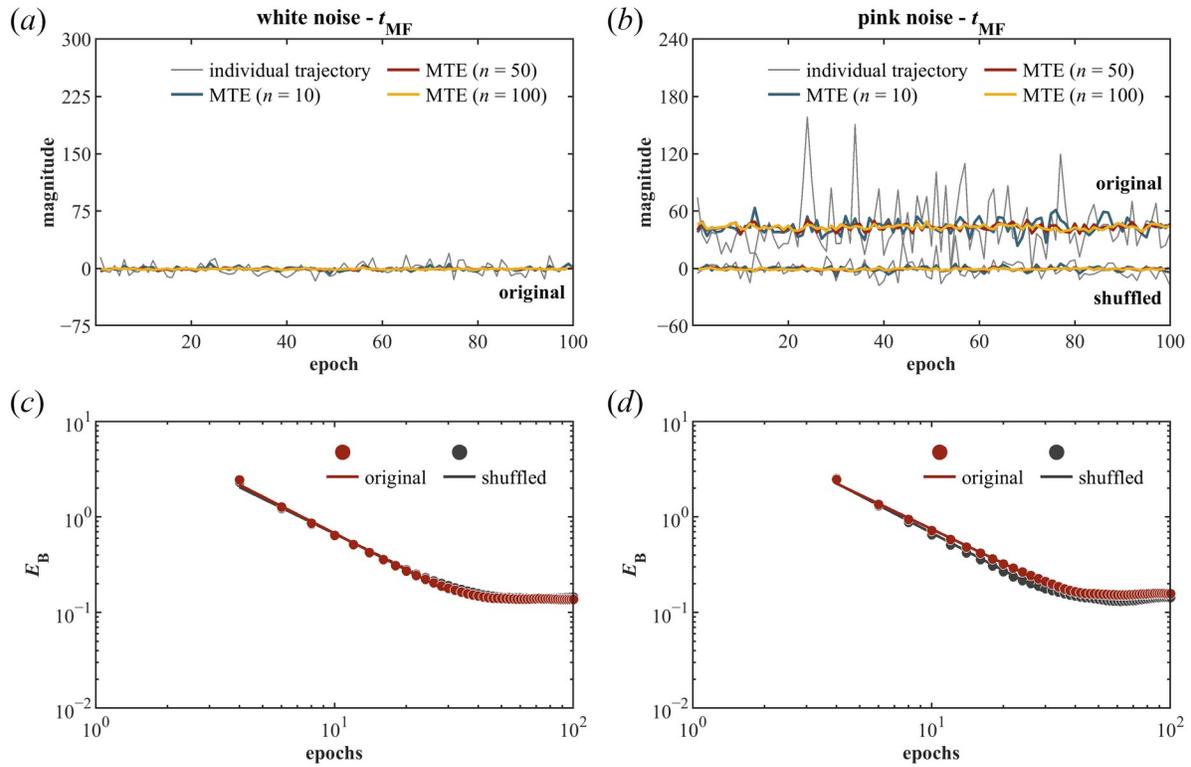

**Figure 8.** Ergodicity breaking in $t_{MF}$ series ($t_{MF}$ calculated across 500-samples epochs). $t_{MF}$ series and MTE average of $t_{MF}$ over 10, 50, and 100 simulations for unsigned white noise (*a*; shuffled series not pictured because of overlap) and unsigned pink noise (*b*). Much like for $H_{fGn}$ series in figures 6*a* and 6*b* and $\Delta a$ series in figures 7*a* and 7*b*, $t_{MF}$ series do not differ between the original and shuffled unsigned white noise, and they take greater values for the original unsigned pink noise than for the original unsigned white noise, showing a consistent differences for larger ensembles. $E_B$ vs. epochs (for lag $\Delta$ = 2 epochs) for (*c*) $t_{MF}$ series for the original and shuffled unsigned white noise and $t_{MF}$ series for the original and shuffled unsigned pinknoise (*d*). These $E_B$ vs. epochs curves for the original and shuffled unsigned series do not overlap with only very small differences in their decay. For instance, the difference between $E_B$ vs. epochs curves for $t_{MF}$ series for the original and shuffled unsigned pink noise is much smaller than for *CV* and *RMS* series in figures 3*d* and 4*d*, respectively. Also, a relatively slower decay in $E_B$ with epochs is observed for $t_{MF}$ series for the original and shuffled unsigned pink noise, but no such distinction is observed for the $t_{MF}$ series for the original and shuffled unsigned white noise.

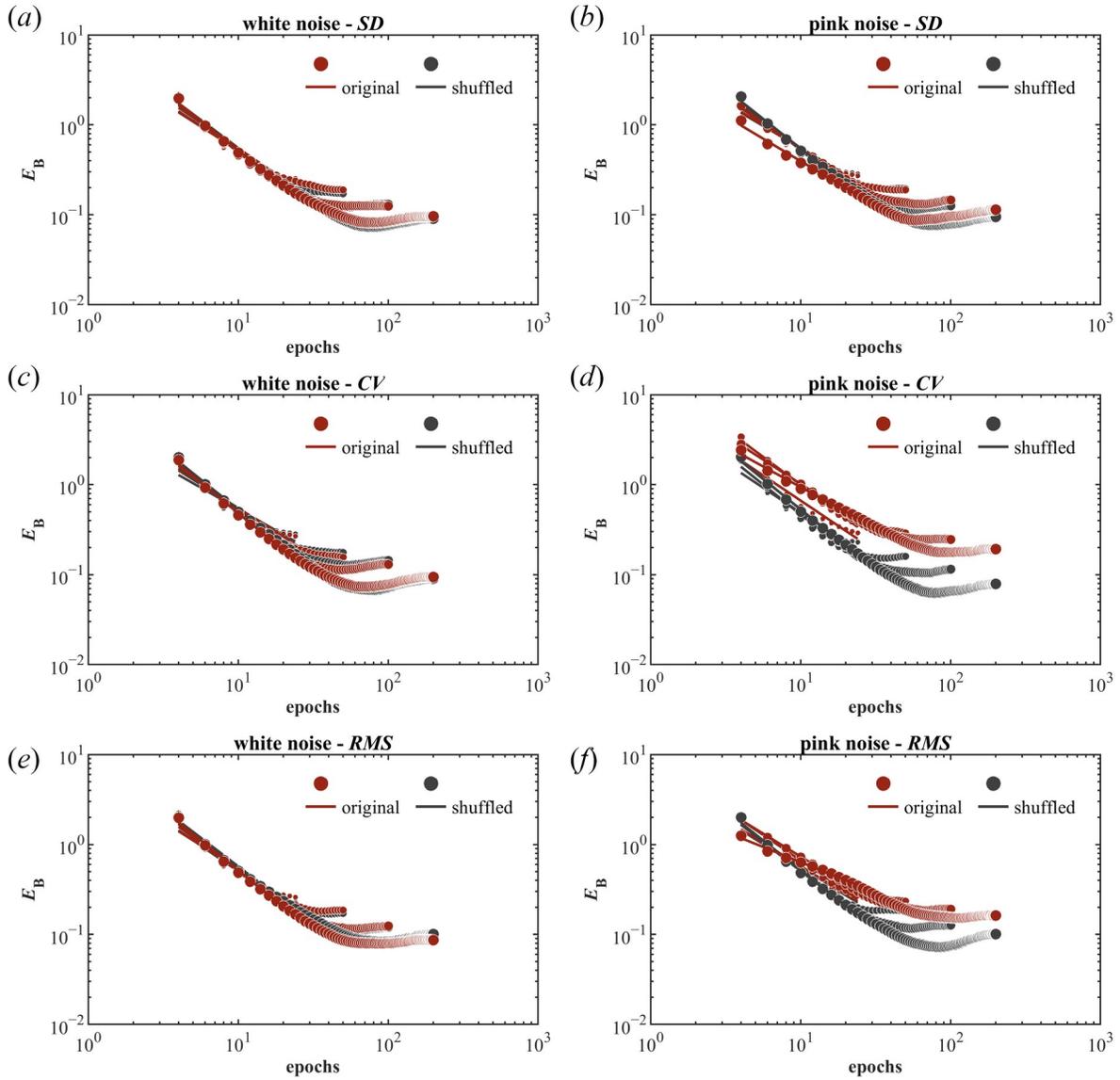

**Figure 9.** $E_B$ vs. epochs curves obtained from the Thirumalai-Mountain method are shown for epochs of different sizes, showing that the results in figures 3 through 5 are largely insensitive to the setting of epoch size. Each panel specifically shows $E_B$ vs. epochs for 25, 50, 100, and 200 epochs of 2000, 1000, 500, and 250 samples each. The unsigned white noise shows no difference from its shuffled counterpart in $E_B$ vs. epochs for (*a*) *SD* series, (*b*) *CV* series, and (*c*) *RMS* series, for different epoch sizes. On the other hand, *CV* and *RMS* series for the original unsigned pink noise show a difference from unsigned *CV* and *RMS* series for the shuffled unsigned pink noise over all epoch sizes (*d* and *f*). As suggested in figure 3, many of the $E_B$ vs. epochs curves show little to no difference in *SD* series for the original and shuffled unsigned pink noise (*b*). However, $E_B$ vs. epochs shows a shallower slope for *SD* series for the original unsigned pink noise than for *SD* series for the shuffled unsigned pink noise for the smallest epochs (i.e., with *s* spanning the longest progression of consecutive epochs). This point suggests that *SD* may behave ergodically but only for relatively large epochs, and it may lose its ergodicity as we use smaller epochs, and in the algorithmic attempt to rescale *SD* by the mean, *CV* and *RMS* series retain the same failure of ergodicity at all epoch sizes.

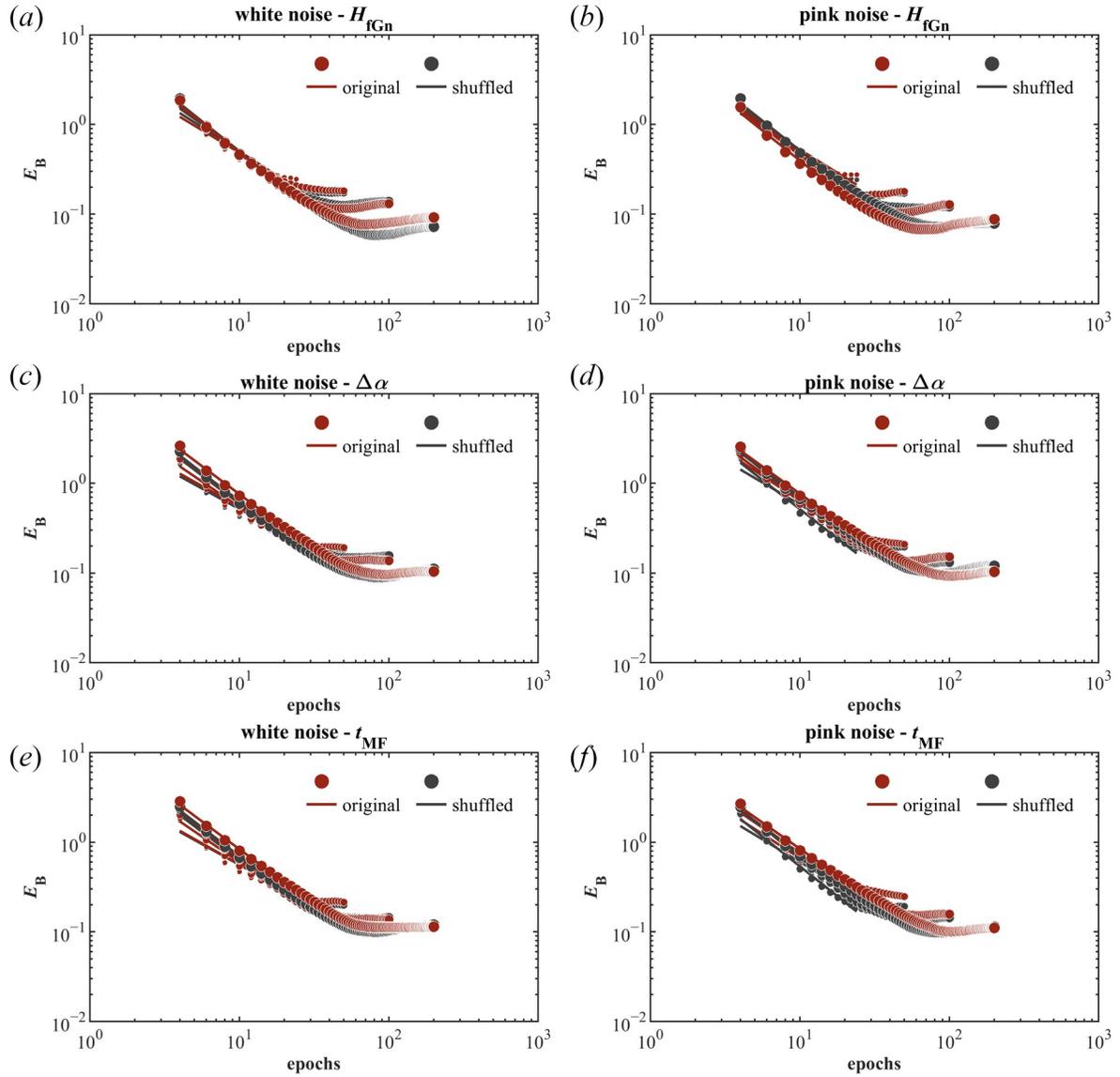

**Figure 10.** $E_B$ vs. epochs curves obtained from the Thirumalai-Mountain method are shown for epochs of different sizes, showing that the results in figures 6 through 8 are largely insensitive to the setting of epoch size. Each panel specifically shows $E_B$ vs. epochs for 25, 50, 100, and 200 epochs of 2000, 1000, 500, and 250 samples each. The original unsigned white noise shows no difference from its shuffled counterpart in $E_B$ vs. epochs for (a) $H_{fGn}$ series, (b) $\Delta a$ series, and (c) $t_{MF}$ series, for different epoch sizes. The original unsigned pink noise largely shows no difference from its shuffled counterpart in $E_B$ vs. epochs for (d) $H_{fGn}$ series, (e) $\Delta a$ series, and (f) $t_{MF}$ series, for different epoch sizes. A minor exception is that the $E_B$ vs. epochs curves for $\Delta a$ and $t_{MF}$ series both become steeper for the original unsigned pink noise, leaving them steeper than and as steep as, respectively, compared to $E_B$ vs epochs curves for their shuffled counterparts.